%% file: paper.tex
\pdfoutput=1
\documentclass[acmsmall,screen,10pt]{acmart}
\renewcommand\footnotetextcopyrightpermission[1]{} %

\usepackage{xspace}
\usepackage{tikz, pgfplots}
\usetikzlibrary{fit, backgrounds, positioning}
\pgfplotsset{compat=1.13}
\pgfplotsset{title style={at={(0.5,0.7)}}}

\begin{document}

\newcommand{\Make}{\textsc{Make}\xspace}
\newcommand{\Rattle}{\textsc{Rattle}\xspace}
\newcommand{\Fabricate}{\textsc{Fabricate}\xspace}
\newcommand{\Bazel}{\textsc{Bazel}\xspace}
\newcommand{\Buck}{\textsc{Buck}\xspace}
\newcommand{\Shake}{\textsc{Shake}\xspace}
\newcommand{\Bigbro}{\textsc{BigBro}\xspace}
\newcommand{\Fac}{\textsc{Fac}\xspace}
\newcommand{\Fsatrace}{\textsc{Fsatrace}\xspace}
\newcommand{\tracedfs}{\textsc{Traced-Fs}\xspace}
\newcommand{\BuildXL}{\textsc{BuildXL}\xspace}
\newcommand{\Nix}{\textsc{Nix}\xspace}
\newcommand{\Memoize}{\textsc{Memoize}\xspace}
\newcommand{\Stroll}{\textsc{Stroll}\xspace}

\newcommand{\postparagraphs}{\vspace{3mm}\noindent}

\title{Build Scripts with Perfect Dependencies}

\author{Sarah Spall}
\affiliation{
  \institution{Indiana University}
}
\author{Neil Mitchell}
\affiliation{
  \institution{Facebook}
}
\author{Sam Tobin-Hochstadt}
\affiliation{
  \institution{Indiana University}
}

\begin{abstract}
Build scripts for most build systems describe the actions to run, and the dependencies between those actions---but often build scripts get those dependencies wrong.
Most build scripts have both \emph{too few dependencies} (leading to incorrect build outputs) and \emph{too many dependencies} (leading to excessive rebuilds and reduced parallelism). Any programmer who has wondered why a small change led to excess compilation, or who resorted to a ``clean'' step, has suffered the ill effects of incorrect dependency specification.
We outline a build system where dependencies are \emph{not specified}, but instead \emph{captured by tracing execution}.
The consequence is that dependencies are always correct by construction and build scripts are easier to write.
The simplest implementation of our approach would lose parallelism, but we are able to recover parallelism using speculation.
\end{abstract}

\maketitle

\input{1-intro}
\input{2-design}
\input{4-implementation}

\input{3-proof}

\input{5-evaluation}
\input{6-related}
\input{7-conclusion}

\bibliography{paper}

\end{document}

%% file: 1-intro.tex
\section{Introduction}
\label{sec:introduction}

Every non-trivial piece of software includes a ``build system'', describing how to set up the system from source code.
Build scripts \cite{build_systems_a_la_carte} describe \emph{commands to run} and \emph{dependencies to respect}. For example, using the \Make build system \cite{make}, a build script might look like:

\vspace{1mm}
\begin{small}
\begin{verbatim}
main.o: main.c
    gcc -c main.c
util.o: util.c
    gcc -c util.c
main.exe: main.o util.o
    gcc -o main.exe main.o util.o
\end{verbatim}
\end{small}
\vspace{1mm}

This script contains three rules. Looking at the first rule, it says \texttt{main.o} depends on \texttt{main.c}, and is produced by running \texttt{gcc -c main.c}. What if we copy the commands into a shell script? We get:

\vspace{1mm}
\begin{small}
\begin{verbatim}
gcc -c main.c
gcc -c util.c
gcc -o main.exe main.o util.o
\end{verbatim}
\end{small}
\vspace{1mm}

That's shorter, simpler and easier to follow. Instead of declaring the outputs and dependencies of each command, we've merely given one valid ordering of the commands (we could equally have put \texttt{gcc -c util.c} first). This simpler specification has additional benefits. First, we've fixed some potential bugs -- these commands depend on the undeclared dependency \texttt{gcc}, and whatever header files are used by \texttt{main.c} and \texttt{util.c}.  These bugs could be fixed in the Makefile by adding a dependency on the version of gcc used, as well as by listing every header file transitively included by \texttt{main.c} and \texttt{util.c}
Furthermore, as the files \texttt{main.c} and \texttt{util.c} evolve, and their dependencies change (by changing the \texttt{\#include} directives), the shell script remains correct, while the \Make script \emph{must} be kept consistent or builds will become incorrect.

\subsection{Why dependencies?}

Given the manifest simplicity of the shell script, why write a \texttt{Makefile}? Build systems such as \Make have two primary advantages, both provided by dependency specification: \textbf{incrementality} and \textbf{parallelism}. \Make is able to re-run only the commands needed when a subset of the files change, saving significant work when only \texttt{util.c} changes. \Make can also run multiple commands in parallel when neither depends on the other, such as the two invocations of \texttt{gcc -c}. With a shell script, these are both challenging research problems in incremental computing and automatic parallelization, respectively, and unlikely to be solvable for arbitrary programs such as \texttt{gcc}.

\subsection{Builds without dependencies}

In this paper we show how to take the above shell script and gain most of the benefits of a \Make build (\S\ref{sec:design}). Firstly, we can skip those commands whose dependencies haven't changed by \emph{tracing} which files they read and write (\S\ref{sec:skipping_unnecessary}) and keeping a history of such traces. Secondly, we can run some commands in parallel, using \emph{speculation} to guess which future commands won't interfere with things already running (\S\ref{sec:speculation}). The key to speculation is a robust model of what ``interfering'' means -- we call a problematic interference a \emph{hazard}, which we define in \S\ref{sec:hazards}.

We have implemented these techniques in a build system called \Rattle\footnote{\url{https://github.com/ndmitchell/rattle}}, introduced in \S\ref{sec:implementation}, which embeds commands in a Haskell script. A key part of the implementation is the ability to trace commands, whose limitations we describe in \S\ref{sec:tracing}. We show that our design decisions produce correct builds in by formalizing hazards and demonstrating important properties about the safety of speculation in \S\ref{sec:proof}. To evaluate our claims, and properly understand the subtleties of our design, we converted existing \Make scripts into \Rattle scripts, and discuss the performance characteristics and \Make script issues uncovered in \S\ref{sec:evaluation}. We also implement two small but non-trivial builds from scratch in \Rattle and report on the lessons learned. Our design can be considered a successor to the \Memoize build system \cite{memoize}, and we compare \Rattle with it and other related work in \S\ref{sec:related}. Finally, in \S\ref{sec:conclusion} we conclude and describe future work.

%% file: 2-design.tex
\section{Build Scripts from Commands}
\label{sec:design}

Our goal is to design a build system where a build script is simply a list of commands. In this section we develop our design, starting with the simplest system that just executes all the commands in order, and ending up with the benefits of a conventional build system.

\subsection{Executing commands}
\label{sec:executing_commands}

Given a build script as a list of commands, like in \S\ref{sec:introduction}, the simplest execution model is to run each command sequentially in the order they were given. Importantly, we require the list of commands is ordered, such that any dependencies are produced before they are used. We consider this sequential execution the reference semantics, and as we develop our design further, require that any optimised/cached implementation gives the same results.

\subsection{Value-dependent commands}
\label{sec:monadic}

While a static list of commands is sufficient for simple builds, it is limited in its expressive power. Taking the build script from \S\ref{sec:introduction}, the user might really want to compile and link \emph{all} \texttt{.c} files -- not just those explicitly listed by the script. A more powerful script might be:

\vspace{3mm}
\begin{small}
\begin{verbatim}
FILES=$(ls *.c)
for FILE in $FILES; do
    gcc -c $FILE
done
gcc -o main.exe ${FILES%
\end{verbatim}
\end{small}
\vspace{3mm}

This script now has a curious mixture of commands (\texttt{ls}, \texttt{gcc}), control logic (\texttt{for}) and simple manipulation (changing file extension\footnote{We take some liberties with shell scripts around replacing extensions, so as not to obscure the main focus.}). Importantly, there is \emph{no fixed list of commands} -- the future commands are determined based on the results of previous commands. Concretely, in this example, the result of \texttt{ls} changes which \texttt{gcc} commands are executed. The transition from a fixed list of commands to a dynamic list matches the \texttt{Applicative} vs \texttt{Monadic} distinction of \citet[\S3.5]{build_systems_a_la_carte}.

There are three approaches to modelling a dynamic list of commands:

\begin{enumerate}
\item We could consider the commands as a stream given to the build system one by one as they are available. The build system has no knowledge of which commands are coming next or how they were created. In this model, a build script supplies a stream of commands, with the invariant that dependencies are produced before they are used, but provides no further information. The main downside is that it becomes impossible to perform any analysis that might guide optimisation.
\item We could expose the full logic of the script to the build system, giving a complete understanding of what commands are computed from previous output, and how that computation is structured. The main downside is that the logic between commands would need to be specified in some constrained domain-specific language (DSL) in order to take advantage of that information. Limiting build scripts to a specific DSL complicates writing such scripts.
\item It would be possible to have a hybrid approach, where dependencies between commands are specified, but the computation is not. Such an approach still complicates specification (some kind of dependency specification is required), but would allow some analysis to be performed.
\end{enumerate}

In order to retain the desired simplicity of shell scripts, we have chosen the first option, modelling a build script as a sequence of commands given to the build system. Future commands may depend on the results of previous commands in ways that are not visible to the build system. The commands are produced with ``cheap'' functions such as control logic and simple manipulations, for example, using a for loop to build a list of object files. We consider the cheap commands to be fixed overhead, run on every build, and not cached or parallelised in any way. If any of these cheap manipulations becomes expensive, they can be replaced by a command, which will then be visible to the build system. The simple list of commands from \S\ref{sec:introduction} is a degenerate case of no interesting logic between commands.

An important consequence of the control logic not being visible to the build system is that the build system has no prior knowledge of which commands are coming next, or if they have changed since it last executed them. As a result, even when the build is a simple static script such as from \S\ref{sec:introduction}, when it is manually edited, the build will execute correctly. The build system is unaware if you edited the script, or if the commands were conditional on something that it cannot observe. Therefore, this model solves the problem of self-tracking from \citet[\S6.5]{build_systems_a_la_carte}.

\subsection{Dependency tracing}
\label{sec:assume_tracing}

For the rest of this section we assume the existence of \emph{dependency tracing} which can tell us which files a command accesses. Concretely, we can run a command in a special mode such that when the command completes (and not before) we can determine which files it read and wrote; these files are considered to be the command's inputs and outputs respectively. We cannot determine at which point during the execution these files were accessed, nor which order they were accessed in. We cannot prevent or otherwise redirect an in-progress access. We discuss the implementation of dependency tracing, and the reasons behind the (frustrating!) limitations, in \S\ref{sec:tracing}.

\subsection{Skipping unnecessary commands}
\label{sec:skipping_unnecessary}

When running a command, we can use system call tracing to capture the files that command reads and writes, and then after the command completes, record the cryptographic hashes of the contents of those files. If the same command is ever run again, and the inputs and outputs haven't changed (have the same hashes), it can be skipped. This approach is the key idea behind both \Memoize\cite{memoize} and \Fabricate\cite{fabricate}. However, this technique makes the assumption that commands are pure functions from their inputs to their outputs, meaning if a command's input files are the same as last time it executed, it will write the same values to the same set of files. Below are four ways that assumption can be violated, along with ways to work around it.

\textbf{Non-deterministic commands} Many commands are non-deterministic -- e.g. the output of \texttt{ghc} object files contains unpredictable values within it (a consequence of the technique described by \citet{lennart:unique_names}). We assume that where such non-determinism exists, any possible output is equally valid. %

\textbf{Incorporating external information} Some commands incorporate system information such as a timestamp, so a cached value will be based on the first time the command was run, not the current time. For compilations that embed the timestamp in metadata, the first timestamp is probably fine. For commands that really want the current time, that step can be lifted into the control logic (as per \S\ref{sec:monadic}) so it will run each time the build runs. Similarly, commands that require unique information, e.g. a GUID or random number, can be moved into control logic and always run.

\textbf{Reading and writing the same file} If a command both reads and writes the same file, and the information written is fundamentally influenced by the file that was read, then the command never reaches a stable state. As an example, \verb"echo x >> foo.txt" will append the character \texttt{x} every time the command is run. Equally, there are also commands that read the existing file to avoid rewriting a file that hasn't changed (e.g. \texttt{ghc} generating a \texttt{.hi} file) and commands that can cheaply update an existing file in some circumstances (the Microsoft C++ linker in incremental mode). We make the assumption that if a command both reads and writes to a file, that the read does not meaningfully influence the write, otherwise it is not really suitable as part of a build system because the command never reaches a stable state and will re-run every time the build is run.

\textbf{Simultaneous modification} If a command reads a file, but before the command completes something else modifies the file (e.g. a human or untracked control logic), then the final hash will not match what the command saw. It is possible to detect such problems with reads by ensuring that the modification time after computing the hash is before the command was started. For simultaneous writes the problem is much harder, so we require that all files produced by the build script are not simultaneously written to.

\postparagraphs In general we assume all commands given to the build system are well behaved and meet the above assumptions.

\subsection{Cloud builds}
\label{sec:cloud_builds}

We can skip execution of a command if all the files accessed have the same hashes as any previous execution (\S\ref{sec:skipping_unnecessary}). However, if \emph{only the files read} match a previous execution, and the files that were written have been stored away, those stored files can be copied over as outputs \emph{without} rerunning the command. If that storage is on a server, multiple users can share the results of one compilation, resulting in cloud build functionality. While this approach works well in theory, there are some problems in practice.

\textbf{Machine-specific outputs} Sometimes a generated output will only be applicable to the machine on which it was generated -- for example if a compiler auto-detects the precise chipset (e.g. presence of AVX2 instructions) or hardcodes machine specific details (e.g. the username). Such information can often be lifted into the command line, e.g. by moving chipset detection into the control logic and passing it explicitly to the compiler. Alternatively, such commands can be explicitly tagged as not being suitable for cloud builds.

\textbf{Relative build directories} Often the current directory, or user's profile directory, will be accessed by commands. These directories change if a user has two working directories, or if they use different machines. We can solve this problem by having a substitution table, replacing values such as the users home directory with \texttt{\$HOME}. If not rectified, this issue reduces the reusability of cloud results, but is otherwise not harmful.

\textbf{Non-deterministic builds} If a command has non-deterministic output, then every time it runs it may generate a different result. Anything that transitively depends on that output is likely to also vary on each run. If the user temporarily disconnects from the shared storage, and runs a non-deterministic command, even if they subsequently reconnect, it is likely anything transitively depending on that command will never match again until after a clean rebuild. There are designs to solve this problem (e.g. the modification comparison mechanism from \citet{erdweg2015sound}), but the issue only reduces the effectiveness of the cloud cache, and usually occurs with intermittent network access, so can often be ignored.

\subsection{Build consistency}
\label{sec:hazards}

As stated in \citet[\S3.6]{build_systems_a_la_carte}, a build is correct provided that:

\begin{quote}
\emph{If we recompute the value of the key (...), we should get exactly the same value as we see in the final store.}
\end{quote}

Specified in terms more applicable to our design, it means that after a build completes, an immediate subsequent rebuild should have no effect because all commands are skipped (assuming the commands given to the build system are the same). However, there are sequences of commands, where each command meets our assumptions separately (as per \S\ref{sec:skipping_unnecessary}), but the combination is problematic:

\vspace{1mm}
\begin{small}
\begin{verbatim}
echo 1 > foo.txt
echo 2 > foo.txt
\end{verbatim}
\end{small}
\vspace{1mm}

\noindent
This program writes \texttt{1} to \texttt{foo.txt}, then writes \texttt{2}. If the commands are re-executed then the first command reruns because its output changed, and after the first command reruns, now the second commands output has changed. More generally, if a build writes different values to the same file multiple times, it is not correct by the above definition, because on a rebuild both commands would re-run. But even without writing to the same file twice, it is possible to have an incorrect build:

\vspace{1mm}
\begin{small}
\begin{verbatim}
sha1sum foo.txt > bar.txt
sha1sum bar.txt > foo.txt
\end{verbatim}
\end{small}
\vspace{1mm}

\noindent
Here \texttt{sha1sum} takes the SHA1 hash of a file, first taking the SHA1 of \texttt{foo.txt} and storing it in \texttt{bar.txt}, then taking the SHA1 of \texttt{bar.txt} and storing it in \texttt{foo.txt}. The problem is that the script first reads from \texttt{foo.txt} on the first line, then writes to \texttt{foo.txt} on the second line, meaning that when the script is rerun the read of \texttt{foo.txt} will have to be repeated as its value has changed.

Writing to a file after it has already been either read or written is the only circumstance in which a build, where every individual command is well-formed (as per \S\ref{sec:skipping_unnecessary}), is incorrect. We define such a build as \emph{hazardous} using the following rules:

\begin{description}
\item[Read then write] If one command reads from a file, and a later command writes to that file, on a future build, the first command will have to be rerun because its input has changed. This behaviour is defined as a \emph{read-write hazard}.  We assume the build author ordered the commands correctly, if not the author can edit the build script.
\item[Write then write] If two commands both write to the same file, on a future build, the first will be rerun (its output has changed), which is likely to then cause the second to be rerun. This behaviour is defined as a \emph{write-write hazard}.
\end{description}

Using tracing we can detect hazards and raise errors if they occur, detecting that a build system is incorrect before unnecessary rebuilds occur. We prove that a build system with deterministic control logic, given the same input the control logic will produce the same output,  and with no hazards always results in no rebuilds in \S\ref{sec:claims}.  The presence of hazards in a build does not guarantee that a rebuild will always occur, for example if the write of a file after it is read does not change the file's value.  But, such a build system is malformed by our definition and if the write definitely can't change the output, then it should not be present.

\subsection{Explicit Parallelism}
\label{sec:explicit_parallelism}

A build script can use explicit parallelism by giving further commands to the build system before previous commands have completed. For example, the script in \S\ref{sec:monadic} has a \texttt{for} loop where the inner commands are independent and could all be given to the build system simultaneously. Such a build system with explicit parallelism must still obey the invariant that the inputs of a command must have been generated before the command is given, requiring some kind of two-way feedback that an enqueued command has completed.

Interestingly, given complete \emph{dependency} information (e.g. as available to \Make) it is possible to infer complete \emph{parallelism} information. However, the difficulty of specifying complete dependency information is the attraction of a tracing based approach to build systems.

\subsection{Implicit Parallelism (Speculation)}
\label{sec:speculation}

While explicit parallelism is useful, it imposes a burden on the build script author. Alternatively we can use implicit parallelism, where some commands are executed speculatively, before they are required by the build script, in the hope that they will be executed by the script in the future and thus skipped by the build system. Importantly, such speculation can be shown to be safe by tracking hazards, provided we introduce a new hazard \emph{speculative-write-read}, corresponding to a speculative command writing to a file that is later read by a command required by the build script (defined precisely in \S\ref{sec:hazards_formally}).

Given a script with no hazards when executed sequentially, we show in \S\ref{sec:claims}: 1) that any ordering of those commands that also has no hazards will result in an equivalent output, see Claim \ref{claim:reorder}; 2) that any parallel or interleaved execution without hazards will also be equivalent, see Claim \ref{claim:parallel}; and 3) if any additional commands are run that don't cause hazards, they can be shown to not change the results the normal build produces, see Claim \ref{claim:additional}. Finally, we prove that if a series of commands contains hazards, so will any execution that includes those required commands, see Claim \ref{claim:keep_hazards}.

As a consequence, if we can predict what commands the build script will execute next, and predict that their execution will not cause hazards, it may be worth speculatively executing them. Effective speculation requires us to predict the following pieces of data.

\textbf{Future commands} The benefit of speculatively executing commands is that they will subsequently be skipped, which only happens if the speculative command indeed occurs in the build script. The simplest way to predict future commands is to assume that they will be the same as they were last time. It is possible to predict in a more nuanced manner given more history of which commands run. Given more information about the build, e.g. all the control logic as per \S\ref{sec:monadic} choice 2, it would be possible to use static analysis to refine the prediction.

\textbf{Absence of hazards} If a hazard occurs the build is no longer correct, and remediation must be taken (e.g. rerunning the build without speculation, see \S\ref{sec:recovering}). Therefore, performance can be significantly diminished if a speculative command leads to a hazard. Given knowledge of the currently running commands, and the files all commands accessed in the last run, it is possible to predict whether a hazard will occur if the access patterns do not change. If tracing made it possible to abort runs that performed hazardous accesses then speculation could be unwound without restarting, but such implementations are difficult (see \S\ref{sec:tracing}).

\textbf{Recovering from Hazards caused by speculation}
If a build using speculative execution causes a hazard, it is possible that the hazard is entirely an artefact of speculation. There are a few actions the build system could take to recover and these are discusses in section \S\ref{sec:proof:classify_hazard}.

%% file: 4-implementation.tex
\section{Implementing \Rattle}
\label{sec:implementation}

We have implemented the design from \S\ref{sec:design} in a build system called \Rattle. We use Haskell as the host language and to write the control logic. Copying the design for \Shake \cite{shake}, a \Rattle build script is a Haskell program that uses the \Rattle library.

\subsection{A \Rattle example}

\begin{figure}
  \begin{small}
\begin{verbatim}
import Development.Rattle
import System.FilePattern
import System.FilePath

main = rattle $ do
    cs <- liftIO $ getDirectoryFiles "." [root </> "*.c"]
    forP cs $ \c -> cmd "gcc -c" c
    cmd "gcc -o main.exe" (map (\x -> takeBaseName x <.> "o") cs)
\end{verbatim}
\end{small}
\caption{A Haskell/\Rattle version of the script from \S\ref{sec:monadic}}
\label{fig:rattle_example}
\end{figure}

\begin{figure}
\begin{small}
\begin{verbatim}
-- The Run monad                           -- Reading/writing files
data Run a = ...                           cmdReadFile :: FilePath -> Run String
rattle :: Run a -> IO a                    cmdWriteFile :: FilePath -> String -> Run ()

-- Running commands                        -- Parallelism
data CmdOption = Cwd FilePath | ...        forP :: [a] -> (a -> Run b) -> Run [b]
cmd :: CmdArguments args => args
\end{verbatim}
\end{small}
\caption{The \Rattle API}
\label{fig:api}
\end{figure}

A complete \Rattle script that compiles all \texttt{.c} files like \S\ref{sec:monadic} is given in Figure \ref{fig:rattle_example}, with the key API functions in Figure \ref{fig:api}. Looking at the example, we see:

\begin{itemize}
\item A \Rattle script is a Haskell program. It makes use of ordinary Haskell imports, and importantly includes \texttt{Development.Rattle}, offering the API from Figure \ref{fig:api}.
\item The \texttt{rattle} function takes a value in the \texttt{Run} monad and executes it in \texttt{IO}. The \texttt{Run} type is the \texttt{IO} monad, enriched with a \texttt{ReaderT} \cite{mtl} containing a reference to shared mutable state (e.g. what commands are in flight, where to store metadata, location of shared storage).
\item All the control logic is in Haskell and can use external libraries -- e.g. \texttt{System.FilePath} for manipulating \texttt{FilePath} values and \texttt{System.FilePattern} for directory listing. Taking the example of replacing the extension from \texttt{.c} to \texttt{.o}, we are able to abstract out this pattern as \texttt{toO} and reuse it later. Arbitrary Haskell IO can be embedded in the script using \texttt{liftIO}. All of the Haskell code (including the IO) is considered control logic and will be repeated in every execution.
\item Commands are given to the build system part of \Rattle using \texttt{cmd}. We have implemented \texttt{cmd} as a variadic function \cite{variadic_functions} which takes a command as a series of \texttt{String} (a series of space-separated arguments), \texttt{[String]} (a list of arguments) and \texttt{CmdOption} (command execution modifiers, e.g. to change the current directory), returning a value of type \texttt{Run ()}. The function \texttt{cmd} only returns once the command has finished executing (whether that is by actual execution, skipping, or fetching from external storage).
\item We have used \texttt{forP} in the example, as opposed to \texttt{forM}, which causes the commands to be given to \Rattle in parallel (\S\ref{sec:explicit_parallelism}). We could have equally used \texttt{forM} and relied on speculation for parallelism (\S\ref{sec:speculation}).
\end{itemize}

Looking at the functions from Figure \ref{fig:api}, there are two functions this example does not use. The \texttt{cmdWriteFile} and \texttt{cmdReadFile} functions are used to perform a read/write of the file system through Haskell code, causing hazards to arise if necessary. Apart from these functions, it is assumed that all Haskell control code only reads and writes files which are not written to by any commands.

\subsection{Alternative \Rattle wrappers}

Given the above API, combined with the choice to treat the control logic as opaque, it is possible to write wrappers that expose \Rattle in new ways. For example, to run a series of commands from a file, we can write:

\vspace{3mm}
\begin{small}
\begin{verbatim}
main = rattle $ do
    [x] <- liftIO getArgs
    src <- readFile x
    forM_ (lines src) cmd
\end{verbatim}
\end{small}
\vspace{3mm}

Here we take a command line argument, read the file it points to, then run each command sequentially using \texttt{forM\_}. We use this script for our evaluation in \S\ref{sec:evaluation}.

An alternative API could be provided by opening up a socket, and allowing a \Rattle server to take command invocations through that socket. Such an API would allow writing \Rattle scripts in other languages, making use of the existing \Rattle implementation. While such an implementation should be easy, we have not yet actually implemented it.

\subsection{Specific design choices and consequences}
\label{sec:choices}

Relative to the reference design in \S\ref{sec:design} we have made a few specific design choices, mostly in the name of implementation simplicity:

\begin{itemize}
\item All of our predictions (see \S\ref{sec:speculation}) only look at the very last run. This approach is simple, and in practice, seems to be sufficient -- most build scripts are run on very similar inputs most of the time.
\item We run a command speculatively if 1) it hasn't been run so far this build; 2) was required in the last run; and 3) doesn't cause a hazard relative to both the completed commands and predicted file accesses of the currently running commands. Importantly, if we are currently running a command we have never run before, we don't speculate anything -- the build system has changed in an unknown way so we take the cautious approach.
\item We never attempt to recover after speculation (see \S\ref{sec:recovering}), simply aborting the build and restarting without any speculation.
\item We treat command lines as black boxes, never examining them to predict their outputs. For many programs a simple analysis (e.g. looking for \texttt{-o} flags) might be predictive.
\item We use a shared drive for sharing build artefacts, but allow the use of tools such as NFS or Samba to provide remote connectivity and thus full ``cloud builds''.
\item \Rattle can go wrong if a speculated command writes to an input file, as per \S\ref{sec:proof:restart_no_speculation}. This problem hasn't occurred in practice, but dividing files into inputs and outputs would be perfectly reasonable. Typically the inputs are either checked into version control or outside the project directory, so that information is readily available.
\item It is important that traces (as stored for \S\ref{sec:skipping_unnecessary}) are only recorded to disk when we can be sure they were not effected by any hazards (see \S\ref{sec:proof:continue}). That determination requires waiting for all commands which ran at the same time as the command in question to have completed.
\item We model queries for information about a file (e.g. existence or modification time) as a read for tracing purposes, thus depending on the contents of the file. For queries about the existence of a file, we rerun if the file contents changes, which may be significantly more frequent than when the file is created or deleted. For queries about modification time, we don't rerun if the modification time changes but the file contents don't, potentially not changing when we should. In practice, most modification time queries are to implement rebuilding logic, so can be safely ignored if the file contents haven't changed.

\end{itemize}

\subsection{Tracing approaches}
\label{sec:tracing}

In \S\ref{sec:assume_tracing} we assume the existence of \emph{dependency tracing} which can, after a command completes, tell us which files that command read and wrote. Unfortunately, such an API is \emph{not} part of the POSIX standard, and is not easily available on any standard platform. We reuse the file access tracing features that have been added to \Shake \cite{neil:file_access}, which in turn are built on top of \Fsatrace \cite{fsatrace}. The techniques and limitations vary by OS:

\begin{itemize}
\item On \textbf{Linux} we use \texttt{LD\_LIBRARY\_PRELOAD} to inject a different C library which records accesses. It can't trace into programs that use system calls directly (typically Go programs \cite{go}) or statically linked binaries. In future we plan to integrate \Bigbro \cite{bigbro} as an alternative, which uses \texttt{ptrace} and fixes these issues.
\item On \textbf{Mac} we also use \texttt{LD\_LIBRARY\_PRELOAD}, but that technique is blocked for security reasons on system binaries, notably the system C/C++ compiler. Installing a C/C++ compiler from another source (e.g. \Nix \cite{nix}) overcomes that limitation.
\item On \textbf{Windows} we can't trace 64bit programs spawned by 32bit programs, but such a pattern is rare (most binaries are now 64bit), often easily remedied (switch to the 64bit version), and possible to fix (although integrating the fix is likely difficult).
\end{itemize}

\noindent In practice, none of the limitations have been overly problematic in the examples we have explored.

We designed \Rattle to work with the limitations of the best cross-platform tracing easily available -- but that involves trade-offs. An enhanced, portable system would be a significant enabler for \Rattle.
Our wishlist for tracing would include a cross-platform API, precise detection times, detection as access happens, and executing code at interception points.

%% file: 3-proof.tex
\renewcommand{\proof}{\vspace{1mm}\noindent \textbf{Proof}: }
\newcommand{\refutation}{\vspace{1mm}\noindent \textbf{Refutation}: }

\section{Correctness}
\label{sec:proof}

\newtheorem{claim}{Claim}

\Rattle's design relies on taking a sequence of commands from the user, and additionally running some commands speculatively. In this section we argue for the correctness of this approach with respect to the reference semantics of sequential evaluation. Throughout, we assume the commands themselves are pure functions from the inputs to their outputs (as reported by tracing), without this property no build system more sophisticated than the shell script of \S\ref{sec:introduction} would be correct.

\subsection{Hazards, formally}
\label{sec:hazards_formally}

In \S\ref{sec:hazards} we introduced the notion of hazards through
intuition and how they are used. Here we give more precise
definitions:

\begin{description}
\item[Read-write hazard] A build has a \emph{read-write hazard} if a file is written to after it has been read. Given the limitations of tracing (see \S\ref{sec:tracing}) we require the stronger property that for all files which are both read and written in a single build, the command which did the write must have finished before any commands reading it start executing, because we cannot assume the command finished writing to the file anytime before it completed.  %
\item[Write-write hazard] A build has a \emph{write-write hazard} if a file is written to after it has already been written. Stated equivalently, a build has a write-write hazard if more than one command in a build writes to the same file.
\item[Speculative-write-read hazard] A build has a \emph{speculative-write-read hazard} if a file is written to by a speculated command, then read from by a command in the build script, before the build script has required the speculated command.
\end{description}

\subsection{Correctness properties}
\label{sec:argument}
\label{sec:claims}

We now state and prove (semi-formally) several key correctness
properties of our approach, particularly in the treatment of hazards. We define the following terms:

\begin{description}
\item[Command] A command $c$ reads from a set of files ($c_r$), and writes to a set of files ($c_w$). We assume the set of files are disjoint ($c_r \cap c_r \equiv \emptyset$), as per \S\ref{sec:skipping_unnecessary}.
\item[Deterministic command] A command is deterministic if for a given set of files and values it reads, it always writes the same values to the same set of files. We assume commands are deterministic.
\item[Fixed point] A deterministic command will have no effect if none of the files it reads or writes have changed since the last time it was run. Since the command is deterministic, and its reads/writes are disjoint, the same reads will take place, the same computation will follow, and then the original values will be rewritten as they were before. We describe such a command as having \emph{reached a fixed point}.
\item[Correct builds] If a build containing a set of commands $C$ has reached a fixed point for every command $c$ within the set, then it is a correct build system as per the definition in \S\ref{sec:hazards}.
\item[No hazards] A build has no hazards if no command in the build writes to a file that a previous command has read or written to.
\item[Input files] We define the input files as those the build reads from but does not write to, namely:
\[
  \bigcup_{c \in C} c_r \setminus \bigcup_{c \in C} c_w
\]
\item[Output files] We define the output files as those the build writes to, namely:
\[
  \bigcup_{c \in C} c_w
\]
\end{description}

\begin{claim}[Completeness] If a set of commands $C$ is run, and no hazards arise, then every command within $C$ has reached a fixed point.
  \label{claim:complete}
\end{claim}

\proof A command $c$ has reached a fixed point if none of the files in $c_r$ or $c_w$ have changed. Taking $c_r$ and $c_w$ separately:

\begin{enumerate}
\item None of the files in $c_w$ have changed. Within a set of commands $C$, if any file was written to by more than one command it would result in a write-write hazard. Therefore, if a command wrote to a file, it must have been the \emph{only} command to write to that file.
\item None of the files $c_r$ have changed. For a read file to have changed would require it to be written to afterwards. If a file is read from then written to it is a read-write hazard, so the files $c_r$ can't have changed.
\end{enumerate}

As a consequence, all commands within $C$ must have reached a fixed point.

\begin{claim}[{Unchanging builds}] A build with deterministic control logic that completes with no hazards has reached a fixed point.  %
\label{claim:no_rebuild}
\end{claim}

\proof The build's deterministic control logic means that beginning with the first command in the build, which is unchanged, given the same set of input files, the command will write to the same set of output files and produce the same results.  It follows that the subsequent command in the build will be unchanged as well since the results of the previous command did not change.  It follows by induction that each of the following commands in the build will remain unchanged.
Because the commands in the build will not have changed, it follows from the proof of Claim \ref{claim:complete} that the commands will not change the values of any files, therefore, the build is at a fixed point.

\begin{claim}[{Reordered builds}]
\label{claim:reorder}

Given a script with no hazards when executed sequentially, with the same initial file contents, any other ordering of those commands that also has no hazards will result in the same terminal file contents.
\end{claim}

\proof The proof of Claim \ref{claim:complete} shows that any script with no hazards will result in a fixed point. We can prove the stronger claim, that for any filesystem state where all \emph{inputs} have identical contents, there is only one fixed point for any ordering of the same commands that has no hazards. For each command, it can only read from files that are in the inputs or files that are the outputs of other commands. Taking these two cases separately:

\begin{enumerate}
\item For read files that are in the inputs, they aren't written to by the build system (or they wouldn't be in the inputs), and they must be the same in both cases (because the filesystem state must be the same for inputs). Therefore, changes to the inputs cannot have an effect on the command.
\item For read files that are the outputs of other commands, they must have been written to \emph{before} this command, or a read-write hazard would be raised.
\end{enumerate}

Therefore, the first command that performs writes cannot access any writes from other commands, and so (assuming determinism) its writes must be a consequence of only inputs. Similarly, the second command can only have accessed inputs and writes produced by the first command, which themselves were consequences of inputs, so that commands writes are also consequences of the input. By induction, we can show that all writes are consequences of the inputs, so identical inputs results in identical writes.

\begin{claim}[{Parallel commands}]
\label{claim:parallel}

Given a script with no hazards when executed sequentially, any parallel or interleaved execution without hazards will also be equivalent.
\end{claim}

\proof None of the proof of Claim \ref{claim:reorder} relies on commands not running in parallel, so that proof still applies.

\begin{claim}[{Additional commands have no effect}]
\label{claim:additional}

 Given a script with no hazards when executed, speculating unnecessary commands that do not lead to hazards will not have an effect on the build's inputs or outputs.
\end{claim}

\proof Provided the inputs are unchanged, and there are no hazards, the proof from Claim \ref{claim:reorder} means the same outputs will be produced. If speculated commands wrote to the outputs, it would result in a write-write hazard. If speculated commands wrote to the inputs before they were used, it would result in a speculative-write-read hazard. If speculated commands wrote to the inputs after they were used, it would result in a read-write hazard. Therefore, assuming no hazards are raised, the speculated commands cannot have an effect on the inputs or outputs.

\begin{claim}[{Preservation of hazards}]
\label{claim:keep_hazards}

 If a sequence of commands leads to a hazard, any additional speculative commands or reordering will still cause a hazard.
\end{claim}

\proof All hazards are based on the observation of file accesses, not the absence of file accesses. Therefore, additional speculative commands do not reduce hazards, only increase them, as additional reads/writes can only cause additional hazards. For reordering, it cannot reduce write-write hazards, which are irrespective of order. For reordering of read-write hazards, it can only remove such a hazard by speculating the writing command before the reading command. However, such speculation is a speculative-read-write hazard, ensuring a hazard will always be raised.

\subsection{Hazard Recovery}
\label{sec:proof:classify_hazard}

If a build involving speculation reaches a hazard there are several remedies that can be taken. In this section we outline these approaches and when each approach is safe to take.

\paragraph{Restart with no speculation}
\label{sec:proof:restart_no_speculation}

Restarting the entire build that happened with speculation, with no speculation, is safe and gives an equivalent result provided the inputs to the build have not changed (Claims \ref{claim:reorder}, \ref{claim:additional}). Unfortunately, while writing to an input is guaranteed to give a hazard, restarting does \emph{not} change the input back. As a concrete example, consider the build script with the single command \texttt{gcc -c main.c}. If we speculate the command \texttt{echo 1 >> main.c} first, it will raise a speculative-write-read hazard, but restarting will not put \texttt{main.c} back to its original value, potentially breaking the build indefinitely. We discuss consequences and possible remediations from speculative commands writing to inputs in \S\ref{sec:choices}, but note such a situation should be very rare and likely easily reversible with minor user intervention.  The build system only executes commands that were previously part of the build script, making it unlikely a build author would include commands that would irreversibly break their build.  For all other recovery actions, we assume that restarting with no speculation is the baseline state, and consider them safe if it is equivalent to restarting with no speculation.
This is the approach we currently take in \Rattle; below we describe
other approaches also justified in our model.

\paragraph{Restart with less speculation}
\label{sec:recovering}

Given the proofs in this section, running again with speculation is still safe. However, running with speculation might repeatedly cause the same speculation hazard, in which case the build would never terminate. Therefore, it is safe to restart with speculation provided there is strictly less speculation, so that (in the worst case) eventually there will be a run with no speculation.

As a practical matter, if a command is part of the commands that cause a hazard, it is a good candidate for excluding from speculation.

\paragraph{Raise a hazard}

Importantly, if there is no speculation (including reordering and parallelism), and a hazard occurs, the hazard can be raised immediately. This property is essential as a base-case for restarting with no speculation.

If two required commands, that is commands in the users build script, cause a write-write hazard, and there have been no speculative-write-read hazards, then the same hazard will be reraised if the command restarts. For write-write, this inevitability is a consequence of Claim \ref{claim:reorder}, as the same writes will occur.
If two required commands cause a read-write hazard, and the relative order they were run in matches their required ordering, then the same hazard will be reraised if the command restarts, because the relative ordering will remain the same, Claim \ref{claim:keep_hazards}. However, if the commands weren't in an order matching their required ordering, it is possible the write will come first avoiding the hazard, so it cannot be immediately raised.

\paragraph{Continue}
\label{sec:proof:continue}

If the hazard only affects speculated commands, and those commands have no subsequent effect on required commands, it is possible to continue the build. However, it is important that the commands have no subsequent effect, even on required commands that have not yet been executed, or even revealed to the build system. Consequently, even if a build can continue after an initial hazard, it may still fail due to a future hazard. A build can continue if the hazards only affect speculated commands, that is either
a \emph{write-write hazard} where both commands were speculated, or
a \emph{read-write hazard} where the read was speculated.
For other hazards, continuing is unsafe. If a speculated command is affected by an initial hazard, and that command is later required by the build script, a new non-continuable hazard will be raised.

%% file: 5-evaluation.tex
\section{Evaluation}
\label{sec:evaluation}

In this section we evaluate the design from \S\ref{sec:design}, specifically our implementation from \S\ref{sec:implementation}. We show how the implementation performs on the example from \S\ref{sec:introduction} in \S\ref{sec:eval:introduction}, on microbenchmarks in \S\ref{sec:eval:overhead}, and then on real projects that currently use \Make{} -- namely FSATrace (\S\ref{sec:eval:fsatrace}), Redis (\S\ref{sec:eval:redis}), Vim (\S\ref{sec:eval:vim}), tmux (\S\ref{sec:eval:tmux}), and Node.js (\S\ref{sec:eval:node}). We implement two sample projects using \Rattle in \S\ref{sec:eval:writing_rattle}.

We have focused our real-world comparison on \Make projects using C because such projects often have minimal external dependencies (making them easier to evaluate in isolation), and despite using the same programming language, have an interesting variety of dependency tracking strategies and size (ranging from one second to three hours) and . Most newer programming languages (e.g. Rust, Haskell etc) have dedicated built tools (e.g. Cargo, Cabal) so small projects in such languages tend to not use general-purpose build tools.

For benchmarks, the first four (\S\ref{sec:eval:introduction}-\S\ref{sec:eval:redis}) were run on a 4 core Intel i7-4790 3.6GHz CPU (16Gb RAM). The remaining benchmarks were run on a 32 core Intel Xeon E7-4830 2.13GHz (64Gb RAM).

\subsection{Validating Rattle's suitability}
\label{sec:eval:introduction}

In \S\ref{sec:introduction} we claimed that the following build script is ``just as good'' as a proper \Make script.

\begin{small}
\begin{verbatim}
gcc -c main.c
gcc -c util.c
gcc -o main.exe main.o util.o
\end{verbatim}
\end{small}

There are two axes on which to measure ``just as good'' -- correctness and performance. Performance can be further broken down into how much rebuilding is avoided, how much parallelism is achieved and how much overhead \Rattle imposes.

\textbf{Correctness} \Rattle is correct, in that the reference semantics is running all the commands, and as we have shown in \S\ref{sec:design} (see  \S\ref{sec:proof} for detailed arguments), and tested for with examples, \Rattle obeys those semantics. In contrast, the \Make version may have missing dependencies which causes it not to rebuild. Examples of failure to rebuild include if there are changes in \texttt{gcc} itself, or changes in any headers used through transitive \texttt{\#include} which are not also listed in the \Make script.

\textbf{Rebuilding too much} \Rattle only reruns a command if some of the files it reads or writes have changed. It is possible that a command only depends on part of a file, but at the level of abstraction \Rattle works, it never rebuilds too much. As a matter of implementation, \Rattle detects changes by hashing the file contents, while \Make uses the modification time. As a consequence, if a file is modified, but its contents do not change (e.g. using \texttt{touch}), \Make will rebuild but \Rattle will not.

\textbf{Parallelism} The script from \S\ref{sec:introduction} has three commands -- the first two can run in parallel, while the the third must wait for both to finish. \Make is given all this information by dependencies, and will always achieve as much parallelism as possible. In contrast, \Rattle has no such knowledge, so it has to recover the parallelism by speculation (see \S\ref{sec:speculation}). During the first execution, \Rattle has no knowledge about even which commands are coming next (as described in \S\ref{sec:monadic}), so it has no choice but to execute each command serially, with less parallelism than \Make. In subsequent executions \Rattle uses speculation to always speculate on the second command (as it never has a hazard with the first), but never speculate on the third until the first two have finished (as they are known to conflict). Interestingly, sometimes \Rattle executes the third command (because it got to that point in the script), and sometimes it speculates it (because the previous two have finished) -- it is a race condition where both alternatives are equivalent. While \Rattle has less parallelism on the first execution, if we were to use shared storage for speculation traces, that can be reduced to the first execution \emph{ever}, rather than the first execution for a given user.

\textbf{Overhead} The overhead inherent in \Rattle is greater than that of \Make, as it hashes files, traces command execution, computes potential hazards, figures out what to speculate and writes to a shared cloud store (in the rest of this section we use a locally mounted shared drive). To measure the overhead, and prove the other claims in this section, we created a very simple pair of files, \texttt{main.c} and \texttt{util.c}, where \texttt{main.c} calls \texttt{printf} using a constant returned by a function in \texttt{util.c}. We then measured the time to do: (1) An initial build from a clean checkout;
(2) a rebuild when nothing had changed;
(3) a rebuild with whitespace changes to \texttt{main.c}, resulting in the same \texttt{main.o} file;
(4) a rebuild with meaningful changes to \texttt{main.c}, resulting in a different \texttt{main.o} file; and
(5) a rebuild with meaningful changes to both C files.

We performed the above steps with 1, 2 and 3 threads, on Linux. To make the parallelism obvious, we modified \texttt{gcc} to sleep for 1 second before starting. The numbers are:

\vspace{2.8mm}
\begin{tabular}{l|r|r||r|r||r|r}
Number of threads & \multicolumn{2}{c||}1 & \multicolumn{2}{c||}2 & \multicolumn{2}{c}3 \\
Tool & \Make & \Rattle & \Make & \Rattle & \Make & \Rattle \\
\hline
1) Initial build & 3.35s & 3.28s & 2.18s & 3.28s & 2.20s & 3.36s \\
2) Nothing changed & 0.00s & 0.00s & 0.00s & 0.00s & 0.00s & 0.00s \\
3) \texttt{main.c} changed whitespace & 2.19s & 1.12s & 2.19s & 1.12s & 2.19s & 1.10s \\
4) \texttt{main.c} changed & 2.19s & 2.16s & 2.23s & 2.21s & 2.19s & 2.22s \\
5) Both C files changed & 3.28s & 3.29s & 2.22s & 2.20s & 2.19s & 2.20s \\
\end{tabular}
\vspace{2.8mm}

As expected, we see that during the initial build \Rattle doesn't exhibit any parallelism, but \Make does (1). In contrast, \Rattle benefits when a file changes in whitespace only and the resulting object file doesn't change, while \Make can't (3). We see 3 threads has no benefit over 2 threads, as this build contains no more parallelism opportunities. Comparing the non-sleep portion of the build, \Make and \Rattle are quite evenly matched, typically within a few milliseconds, showing low overheads. We focus on the overheads in the next section.

\subsection{Measuring overhead}
\label{sec:eval:overhead}

In order to determine what overhead \Rattle introduces, we ran a fixed set of commands with increasingly more parts of \Rattle enabled. \Rattle command execution builds on the command execution from \Shake \cite{shake}, which uses \Fsatrace for tracing and the Haskell \texttt{process} library for command execution. We ran the commands in a clean build directory in 7 ways:

\begin{enumerate}
\item Using \texttt{make -j1} (\Make with 1 thread), as a baseline.
\item Using \texttt{System.Process} from the Haskell \texttt{process} library.
\item Using \texttt{cmd} from the  \Shake library \cite{shake}, which builds on  the \texttt{process} library.
\item Using \texttt{cmd} from \Shake, but wrapping the command with \Fsatrace for file tracing.
\item Using \texttt{cmd} from \Shake with the \texttt{Traced} setting, which runs \Fsatrace and collects the results.
\item Using \Rattle with no speculation or parallelism, and without shared storage.
\item Using \Rattle with all features, including shared storage.
\end{enumerate}

To obtain a set of commands typical of building, we took the latest version of \Fsatrace\footnote{\url{https://github.com/jacereda/fsatrace/commit/41fbba17da580f81ababb32ec7e6e5fd49f11473}} and ran \texttt{make -j1} with maximum verbosity,  recording the commands that were printed to the standard output. On Windows \Fsatrace runs 25 commands (21 compiles, 4 links). On Linux \Fsatrace runs 9 commands (7 compiles, 2 links). On Linux the list of commands produces write-write hazards, because it compiles some files (e.g. \texttt{shm.c}) twice, once with \texttt{-fPIC} (position independent code), and once without. However, both times it passes \texttt{-MMD} to cause \texttt{gcc} to produce \texttt{shm.d} at the same location (\texttt{shm.d} is used to track dependencies). That write-write hazard is a genuine problem, and in an incredibly unlucky build \Fsatrace would end up with a corrupted \texttt{shm.d} file. To remedy the problem we removed the \texttt{-MMD} flag as it doesn't impact the benchmark and isn't required by \Rattle. We ran all sets of commands on both Windows and Linux.

\vspace{2.8mm}
\begin{tabular}{l|rrr|rrr}
Commands & \multicolumn{3}{c|}{Windows} & \multicolumn{3}{c}{Linux} \\
\hline
1) Make                      &  9.96s & 100\% &       &    1.19s & 100\% & \\
2) process                   & 10.26s & 103\% &  +3\% &    1.18s &  99\% & -1\% \\
3) \Shake                    & 10.58s & 106\% &  +3\% &    1.17s &  98\% & -1\% \\
4) \Shake + \Fsatrace        & 12.66s & 127\% & +21\% &    1.23s & 103\% & +5\% \\
5) \Shake + \texttt{Traced}  & 13.06s & 131\% &  +4\% &    1.23s & 103\% & +0\% \\
6) \Rattle                   & 14.43s & 145\% & +14\% &    1.25s & 105\% & +2\% \\
7) \Rattle + everything      & 14.53s & 146\% &  +1\% &    1.27s & 107\% & +2\% \\
\end{tabular}
\vspace{2.8mm}

In the above chart both Windows and Linux have three columns -- the time taken (average of the three fastest of five runs), that time as a percentage of the \Make run, and the delta from the row above. The results are significantly different between platforms:

On Windows, we see that the total overhead of \Rattle makes the execution 46\% slower. 21\% of the slowdown is from \Fsatrace (due to hooking Windows kernel API), with the next greatest overhead being from \Rattle itself. Of the \Rattle overhead, the greatest slowdown is caused by canonicalizing paths. Using the default NTFS file system, Windows considers paths to be case insensitive. As a result, we observe paths like \verb"C:\windows\system32\KERNELBASE.dll", which on disk are called \verb"C:\Windows\System32\KernelBase.dll", but can also be accessed by names such as \verb"C:\Windows\System32\KERNEL~1.DLL". Unfortunately, Windows also supports case sensitive file systems, so  using case-insensitive equality is insufficient.
On Windows, enabling the default anti-virus (Windows Defender) has a significant impact on the result, increasing the \Make baseline by 11\% and the final time by 29\%. These results were collected with the anti-virus disabled.

On Linux, the total overhead is only 7\%, of which nearly all (5\%) comes from tracing.

These results show that tracing has a minor but not insignificant effect on Linux, whereas on Windows can be a substantial performance reduction. Additionally, the Windows results had far more variance, including large outliers. Thus, our remaining benchmarks were all run on Linux.

\subsection{\Fsatrace}
\label{sec:eval:fsatrace}

To compare \Make and \Rattle building \Fsatrace we took the commands we extracted for \S\ref{sec:eval:overhead} and ran the build script for the 100 previous commits in turn, starting with a clean build then performing incremental builds. To make the results readable, we hid any commits where all builds took < 0.02s, resulting in 26 interesting commits. We ran with 1 to 4 threads, but omit the 2 and 3 thread case as they typically fall either on or just above the 4 thread results.

\begin{tikzpicture}
\begin{axis}[
  title={Compile time at each successive commit\hspace{2cm}\phantom{.}},
  width=\textwidth,
  height=4cm,
  ylabel={Seconds},
  tick pos=left,
  ymin=0,
  xmin=0,
  xmax=16,
  axis x line*=bottom,
  axis y line*=left,
  legend style={draw=none},
]
\addplot [color=cyan, mark=o] table [x expr=\coordindex, y=make1] {data/fsatrace.dat};    \addlegendentry{\Make 1 thread}
\addplot [color=cyan, mark=*] table [x expr=\coordindex, y=make4] {data/fsatrace.dat};    \addlegendentry{\Make 4 threads}
\addplot [color=purple, mark=triangle] table [x expr=\coordindex, y=rattle1] {data/fsatrace.dat};  \addlegendentry{\Rattle 1 thread}
\addplot [color=purple, mark=triangle*] table [x expr=\coordindex, y=rattle4] {data/fsatrace.dat};  \addlegendentry{\Rattle 4 threads}
\end{axis}
\end{tikzpicture}

As we can see, the first build is always > 1s for \Rattle, but \Make is able to optimise it as low as 0.33s with 4 threads. Otherwise, \Rattle and \Make are competitive -- users would struggle to see the difference. The one commit that does show some variation is commit 2, where both are similar for 1 thread, but \Rattle at 4 threads is slightly slower than 1 thread, while \Make is faster. The cause is speculation leading to a write-write hazard. Concretely, the command for linking \texttt{fsatrace.so} changed to include a new file \texttt{proc.o}. \Rattle starts speculating on the old link command, then gets the command for the new link. Both commands write to \texttt{fsatrace.so}, leading to a hazard, and causing \Rattle to restart without speculation.

\subsection{Redis}
\label{sec:eval:redis}

\begin{tikzpicture}
\begin{axis}[
  title={Compile time at each successive commit\hspace{2cm}\phantom{.}},
  width=\textwidth,
  height=4cm,
  ylabel={Seconds},
  tick pos=left,
  ymin=0,
  xmin=0,
  axis x line*=bottom,
  axis y line*=left,
  legend style={draw=none},
  xmax=34,
]
\addplot [color=cyan, mark=o] table [x expr=\coordindex, y=make1] {data/redis.dat}; 	\addlegendentry{\Make 1 thread}
\addplot [color=cyan, mark=*] table [x expr=\coordindex, y=make4] {data/redis.dat}; 	\addlegendentry{\Make 4 threads}
\addplot [color=purple, mark=triangle] table [x expr=\coordindex, y=rattle1] {data/redis.dat}; 	\addlegendentry{\Rattle 1 thread}
\addplot [color=purple, mark=triangle*] table [x expr=\coordindex, y=rattle4] {data/redis.dat}; 	\addlegendentry{\Rattle 4 threads}
\end{axis}
\end{tikzpicture}

Redis is an in-memory NoSQL database written in C, developed over the last 10 years. Redis is built using recursive \Make \cite{miller:recursive_make}, where the external dependencies of Redis (e.g. Lua and Linenoise) have their own \Make scripts which are called by the main Redis \Make script. Using stamp files, on successive builds, the external dependencies are not checked -- meaning that users must clean their build when external dependencies change.

When benchmarking \Rattle vs \Make we found that \Rattle was about 10\% slower due to copying files to the cloud cache (see \S\ref{sec:cloud_builds}), something \Make does not do. Therefore, for a fairer comparison, we disabled the cloud cache. For local builds, the consequence of disabling the cloud cache is that if a file changes, then changes back to a previous state, we will have to rebuild rather than get a cache hit -- something that never happens in this benchmark. With those changes \Rattle is able to perform very similarly to \Make, even while properly checking dependencies. Differences occur on the initial build, where \Rattle has no information to speculate with, and commit 1, where \Rattle correctly detects no work is required but \Make rebuilds some dependency information.

\subsection{Vim}
\label{sec:eval:vim}

\begin{tikzpicture}
\begin{axis}[
  title={Compile time at each successive commit\hspace{2cm}\phantom{.}},
  width=\textwidth,
  height=3.7cm,
  ylabel={Seconds},
  tick pos=left,
  ymin=0,
  axis x line*=bottom,
  axis y line*=left,
  legend style={draw=none},
  xmin=0,
  xmax=43,
]

\addplot [color=cyan, mark=o] table [x expr=\coordindex, y=make1] {data/vim.dat}; 	\addlegendentry{\Make 1 thread}
\addplot [color=cyan, mark=*] table [x expr=\coordindex, y=make4] {data/vim.dat}; 	\addlegendentry{\Make 4 threads}
\addplot [color=purple, mark=triangle] table [x expr=\coordindex, y=rattle1_noshared] {data/vim.dat}; 	\addlegendentry{\Rattle 1 thread}
\addplot [color=purple, mark=triangle*] table [x expr=\coordindex, y=rattle4_noshared] {data/vim.dat}; 	\addlegendentry{\Rattle 4 threads}
\end{axis}
\end{tikzpicture}

Vim is a popular text editor, over 25 years old. The majority of the source code is Vim script and C, and it is built with \Make. To compare the original \Make based build to \Rattle, we built Vim over a 35 commits, using both the original \Make and a \Rattle script created by recording the commands \Make executed on a clean build. The benchmarks show that \Make and \Rattle had similar performance for almost every commit, apart from two.

\begin{enumerate}
\item For the 20th commit in the chart, \Make did much more work than \Rattle. This commit changed the script which generates \texttt{osdef.h}, in such a way that it did not change the contents of \texttt{osdef.h}. \Make rebuilt all of the object files which depend on this header file, while \Rattle skipped them as the file contents had not changed.
\item For the 29th commit in the chart, \Rattle did much more work than \Make. This commit changed a prototype file that is indirectly included by all object files.  This prototype file is not listed as a dependency of the object files, so \Make did not rebuild them, but \Rattle noticed the changed dependency.  This missing dependency could have resulted in an incorrect build or a build error, but in this case (by luck) seems not to have been significant. A future prototype change may result in incorrect builds.
\end{enumerate}

\subsection{tmux}
\label{sec:eval:tmux}

\begin{tikzpicture}
\begin{axis}[
  title={Compile time at each successive commit\hspace{2cm}\phantom{.}},
  width=\textwidth,
  height=3.7cm,
  ylabel={Seconds},
  ymax=50,
  ymin=0,
  xmin=0,
  axis x line*=bottom,
  axis y line*=left,
  legend style={draw=none},
  tick pos=left
]

\addplot [color=cyan, mark=o] table [x expr=\coordindex, y=make1] {data/tmux.dat}; 	\addlegendentry{\Make 1 thread}
\addplot [color=cyan, mark=*] table [x expr=\coordindex, y=make4] {data/tmux.dat}; 	\addlegendentry{\Make 4 threads}
\addplot [color=purple, mark=triangle] table [x expr=\coordindex, y=rattle1_noshared] {data/tmux.dat}; 	\addlegendentry{\Rattle 1 thread}
\addplot [color=purple, mark=triangle*] table [x expr=\coordindex, y=rattle4_noshared] {data/tmux.dat}; 	\addlegendentry{\Rattle 4 threads}
\end{axis}
\end{tikzpicture}

tmux is a terminal multiplexer in development for the last 12 years.  It is built from source using a combination of a conventional \verb|sh autogen.sh && ./configure && make| sequence, making use of \textsc{Autoconf}, \textsc{Automake}, \textsc{pkg-config}, and \Make.
To compare a \Rattle based tmux build to the \Make based one, we first ran \texttt{sh autogen.sh \&\& ./configure}, and then used the resulting \Make scripts to generate \Rattle build scripts. tmux was built over a series of 38 commits with both \Make and \Rattle. For the 11th, 22nd and 23rd commits \Rattle and \Make took noticeably different amounts of time. In all cases the cause was a build script change, preventing \Rattle from speculating, and thus taking approximately as long as a full single-threaded build.  For those three commits where the build script changed, it was possible for a write-write hazard to occur, and we occasionally observed it for the 11th and 22nd commits (but not in the 5 runs measured above).

\subsection{Node.js}
\label{sec:eval:node}

Node.js is a JavaScript runtime built on Chrome's V8 Javascript engine that has been in development since 2009.  The project is largely written in JavaScript, C++, Python and C, and is built using \Make and a meta-build tool called Generate Your Projects (GYP) \cite{gyp}.

\paragraph{Node.js build system}

To build Node.js from source, a user first runs \texttt{./configure}, which configures the build and runs GYP, generating Makefiles. From then on, a user can build with \Make. GYP generates the majority of the Makefiles used to build Node.js, starting from a series of source \texttt{.gyp} files specifying metadata such as targets and dependencies. GYP generates a separate \texttt{*.mk} Makefile for each target, all of which are included by a generated top-level Makefile. That top-level Makefile can be run by \Make, but through a number of tricks, GYP  controls  the dependency calculation:

\begin{enumerate}
\item All tracked commands are run using the \texttt{do\_cmd} function from the top-level Makefile. The \texttt{do\_cmd} function checks the command, and if anything has changed since the last run, reruns it. To ensure this function is called on every run, targets whose rules use \texttt{do\_cmd} have a fake dependency on the \texttt{phony} target \texttt{FORCE\_DO\_CMD}, forcing \Make to rerun them every time. This mechanism operates much like \Rattle command skipping from \S\ref{sec:skipping_unnecessary}.
\item Tracked commands create dependency files for each target, recording observed dependencies from the last run. Typically, this information is produced by the compiler, e.g. \texttt{gcc -MMD}.
\item Generated source files are produced and managed using \texttt{.intermediate} targets. By relying on a combination of \texttt{do\_cmd}, \texttt{FORCE\_DO\_CMD} and a special prerequisite \texttt{.INTERMEDIATE} that depends on all \texttt{.intermediate} targets, the intention appears to be to regenerate the generated code whenever something relevant has changed. However, in our experiments, those intermediate targets seem to run every time.
\end{enumerate}

It appears the build of Node.js has tried to take an approach closer to that of \Rattle, namely, tracking dependencies each time a command runs and making sure commands re-run when they have changed.  The implementation is quite complicated and relies on deep knowledge of precisely how \Make operates, but still seems to not match the intention.

%
%

%

%
%
%
%
%
%
%

%
%

%

  %

%

%

\vspace{3mm}
\begin{tikzpicture}
\begin{axis}[ymode=log,
  title={Compile time at each successive commit\hspace{2cm}\phantom{.}},
  width=\textwidth,
  tick pos=left,
  height=4.5cm,
  ylabel={Seconds (log scale)},
  xmin=0,
  axis x line*=bottom,
  axis y line*=left,
  legend style={draw=none},
]

\addplot [color=cyan, mark=o] table [x expr=\coordindex, y=make1] {data/node.dat}; 	\addlegendentry{\Make 1 thread}
\addplot [color=cyan, mark=*] table [x expr=\coordindex, y=make4] {data/node.dat}; 	\addlegendentry{\Make 4 threads}
\addplot [color=purple, mark=triangle] table [x expr=\coordindex, y=rattle1_noshared] {data/node.dat}; 	\addlegendentry{\Rattle 1 thread}
\addplot [color=purple, mark=triangle*] table [x expr=\coordindex, y=rattle4_noshared] {data/node.dat}; 	\addlegendentry{\Rattle 4 threads}
\end{axis}
\end{tikzpicture}

To compare to \Rattle, we built Node.js over a series of 38 commits with both \Make and \Rattle, with 1 thread and 4 threads. The \Rattle build was created for each commit by recording every command executed by the original build, excluding the commands exclusively dealing with dependency files. Due to the large variations in time, the above plot uses a log scale.

For building from scratch, \Make and \Rattle took approximately the same amount of time with a single thread. However, with 4 threads \Make was able to build Node nearly four times faster, since \Rattle had no speculation information. For many commits \Rattle and \Make give similar results, but for those with the least amount of work, where only a few files change, \Rattle is able to beat \Make. The fastest observed \Make time is 35 seconds, while \Rattle goes as low as 12 seconds. The major cause is that \Make \emph{always} rebuilds the generated files but \Rattle (correctly) does not.

For three commits, 2, 12, and 29, the \Rattle build with 4 threads was slower than single threaded \Rattle.  The cause was read-write hazards occurring, forcing \Rattle to restart the build with no speculation.  The build script changed at each of these commits, and so \Rattle speculating commands from the previous script led to hazards with the changed commands in the new script.

\subsection{Summary comparing with \Make}

Looking across the projects in this section, a few themes emerge.

\begin{itemize}
\item For most commits, \Rattle and \Make give similar performance. The one case that is not true is the initial build at higher levels of parallelism, where \Make outperforms \Rattle, because \Rattle has no speculation information. If \Rattle used a shared speculation cache for the initial build that difference would disappear. Such a cache would only need to provide approximate information to get most of the benefits, so the nearest ancestor commit with such information should be sufficient -- there would be no need to maintain or curate the list.
\item The difference between 1 thread and 4 threads can be quite significant, and typically \Make and \Rattle do equally well with parallelism. The implicit parallelism in \Rattle is an important feature which works as designed.
\item In rare cases \Rattle suffers a hazard, which can be observed on the build time plot.
\item Every \Make project has a minor variation on how dependencies are managed. Most make use of \texttt{gcc} to generate dependency information%
  . Some provide it directly (e.g. \Fsatrace), some also use stamp files (e.g. Redis) and some build their own mechanisms on top of \Make (e.g. Node.js, but also \citet[\S2]{hadrian}).
\item Many projects give up on dependencies in less common cases, requiring user action, e.g. Redis requires a clean build if a dependency changes.
\item \Rattle correctly rebuilds if the build file itself updates, but most of the \Make projects do not. The one exception is Node.js, through its use of its dependency encoding in the Makefile.
\end{itemize}

\subsection{Writing \Rattle scripts}
\label{sec:eval:writing_rattle}

In order to see how difficult it is to write build scripts for \Rattle, we tried writing two examples -- namely \Fsatrace, and a program to fetch Haskell dependencies and install them.

\paragraph{Building \Fsatrace}

We implemented a \Rattle script to build \Fsatrace. Unlike \S\ref{sec:eval:fsatrace}, where we took the outputs from \Make, this time we took the source Makefiles and tried to port their logic to \Rattle. We did not port the testing or \texttt{clean} phony target, but we did port both Windows and Linux support. The script first defines 11 variables, following much the same pattern as the Makefile defines the same variables -- some of which are conditional on the OS. Next we compile all the C sources, and finally we do a few links. The total code is 19 non-comment lines, corresponding to 46 non-comment lines in the original Makefiles.

Given the setup of the code, it was trivial to use explicit parallelism for compiling all C files, but doing so for the linking steps, and arranging for the links to be performed as soon as possible, would have been tedious. Taking some representative pieces from the build, we have:

\begin{small}
\begin{verbatim}
let cppflags | isWindows = "-D_WIN32_WINNT=0x600 -isysteminclude/ddk"
             | otherwise = "-D_GNU_SOURCE -D_DEFAULT_SOURCE=1"
let sosrcs = ["src/unix/fsatraceso.c","src/unix/shm.c","src/unix/proc.c"]

forP_ (srcs ++ if isWindows then dllsrcs else sosrcs) $ \x ->
    cmd "gcc -c" ["-fPIC" | not isWindows] cppflags cflags x "-o" (x -<.> "o")

cmd "gcc" ldflags ldobjs (map (-<.> "o") srcs) ldlibs "-o" ("fsatrace" <.> exe)
\end{verbatim}
\end{small}

The overall experience of porting the Makefile was pleasant. The explicit \texttt{isWindows} is much nicer than the Makefile approach. The  availability of lists and a rich set of operations on them made certain cases more natural. The use of functions, e.g. \texttt{objects}, allowed simplifying transformations and giving more semantically meaningful names. The powerful syntax of \texttt{cmd}, allowing multiple arguments, makes the code look relatively simple. The only downside compared to \Make was the use of quotes in more places.

\paragraph{Haskell Dependencies}

As a second test, we wrote a program to fetch and build a Haskell library and its dependencies. For Haskell packages, Cabal is both a low-level library/interface for configuring, building and installing, and also a high-level tool for building sets of packages. Stackage is a set of packages which are known to work together. Stack is a tool that consumes the list of Stackage packages, and using low-level Cabal commands, tries to install them. We decided to try and replicate Stack, with the original motivation that by building on top of \Rattle we could get cloud builds of Haskell libraries for free.

We were able to write a prototype of something Stack-like using \Rattle in 71 non-comment lines. Our prototype is capable of building some projects, showing the fundamental mechanisms work, but due to the number of corner cases in Haskell packages, is unlikely to be practical without further development. The main logic that drives the process is reproduced below.

\begin{small}
\begin{verbatim}
stack :: String -> [PackageName] -> Run ()
stack resolver packages = do
    let url = "https://www.stackage.org/" ++ resolver ++ "/cabal.config"
    cmd Shell "curl -sSL" url "-o" resolver <.> "config"
    versions <- liftIO $ readResolver $ resolver <.> "config"
    flags <- liftIO extraConfigureFlags
    needPkg <- memoRec $ \needPkg name -> do
        let v = version Map.! name
        whenJust v $ installPackage needPkg config flags name
        return v
    forP_ packages needPkg
\end{verbatim}
\end{small}

Reading this code, given a Stackage resolver (e.g. \texttt{nightly-2012-10-29}), we download the resolver to disk, read it, then recursively build projects in an order satisfying dependencies. The download uses \texttt{curl}, which has no meaningful reads, so will always be skipped. Importantly, the URL includes the full resolver which will never change once published. The parsing of the resolver is Haskell code, not visible to \Rattle, repeated on every run (the time taken is negligible). We then build a recursive memoisation pattern using \texttt{memoRec}, where each package is only built once, and while building a package you can recursively ask for other packages. We then finally demand the packages that were requested. %

Writing the program itself was complex, due to Haskell packaging corner cases -- but the \Rattle aspect was pleasant. Having Haskell available to parse the configuration and having efficient \texttt{Map} data structures made it feasible. Not worrying about dependencies, focusing on it working, vastly simplified development. There were lots of complications due to the nature of Haskell packages (e.g. some tools/packages are special, parsing metadata is harder than it should be), but only a few which interacted closely with \Rattle:

\begin{enumerate}
\item There are some lock files, e.g. \texttt{package.cache.lock}, which are both read and written by many commands, but not in an interesting way. We explicitly ignored these files from being considered for hazards.
\item Haskell packages are installed in a database, which by default is appended to with each package registration. That pattern relies on the order of packages and is problematic for concurrency. Many build systems, e.g. \Bazel, use multiple package databases, each containing one package. We follow the same trick.
\item By default, Cabal produces packages where some fields (but not all) include an absolute path. To make the results generally applicable for cloud builds, we remove the absolute paths. To ensure the absolute paths never make it to the cloud, we run the command that produces and the command that cleans up as a single command from the point of view of \Rattle.
\item Since cachable things have to be commands, that required putting logic into commands, not just shell calls. To make that pleasant, we wrote a small utility to take Template Haskell commands \cite{template_haskell} and turn them into Haskell scripts that could then be run from the command line.
\end{enumerate}

%% file: 6-related.tex
\section{Related work}
\label{sec:related}

The vast majority of existing build systems are \emph{backward build systems} -- they start at the final target, and recursively determine the dependencies required for that target. In contrast, \Rattle is a \emph{forward build system}---the script executes sequentially in the order given by the user.

\subsection{Comparison to forward build systems}

The idea of treating a script as a build system, omitting commands that have not changed, was pioneered by \Memoize \cite{memoize} and popularised by \Fabricate \cite{fabricate}. In both cases the build script was written in Python, where cheap logic was specified in Python and commands were run in a traced environment. If a traced command hadn't changed inputs  since it was last run, then it was skipped. We focus on \Fabricate, which came later and offered more features. \Fabricate uses \texttt{strace} on Linux, and on Windows uses either file access times (which are either disabled or very low resolution on modern Windows installations) or a proprietary and unavailable \texttt{tracker} program. Parallelism can be annotated explicitly, but often omitted.

These systems did not seem to have much adoption -- in both cases the original sources are no longer available, and knowledge of them survives only as GitHub copies. \Rattle differs from an engineering perspective by the tracing mechanisms available (see \S\ref{sec:tracing}) and the availability of cloud build (see \S\ref{sec:cloud_builds}) -- both of which are likely just a consequence of being developed a decade later. \Rattle extends these systems in a more fundamental way with the notion of hazards, which both allows the detection of bad build scripts, and allows for speculation -- overcoming the main disadvantage of earlier forward build systems. Stated alternatively, \Rattle takes the delightfully simple approach of these build systems, and tries a more sophisticated execution strategy.

Recently there have been three other implementations of forward build systems we are aware of.

\begin{enumerate}
\item \Shake \cite{shake} provides a forward mode implemented in terms of a backwards build system. The approach is similar to the \Fabricate design, offering skipping of repeated commands and explicit parallelism. In addition, \Shake allows caching custom functions as though they were commands, relying on the explicit dependency tracking functions such as \texttt{need} already built into \Shake. The forward mode has been adopted by a few projects, notably a library for generating static websites/blogs. The tracing features in \Rattle are actually provided by \Shake as a library (which in turn builds on \Fsatrace).
\item \Fac \cite{fac} is based on the \Bigbro tracing library. Commands are given in a custom file format as a static list of commands (i.e. no monadic expressive power as per \S\ref{sec:monadic}), but may optionally include a subset of their inputs or outputs. The commands are not given in any order, but the specified inputs/outputs are used to form a dependency order which \Fac uses. If the specified inputs/outputs are insufficient to give a working order, then \Fac will fail but record the \emph{actual} dependencies which will be used next time -- a build with no dependencies can usually be made to work by running \Fac multiple times.
\item \Stroll \cite{stroll} takes a static set of commands, without either a valid sequence or any input/output information, and keeps running commands until they have all succeeded. As a consequence, \Stroll may run the same command multiple times, using tracing to figure out what might have changed to turn a previous failure into a future success. \Stroll also reuses the tracing features of \Shake and \Fsatrace.
\end{enumerate}

There are significantly fewer forward build systems than backwards build systems, but the interesting dimension starting to emerge is how an ordering is specified. The three current alternatives are the user specifies a valid order (\Fabricate and \Rattle), the user specifies partial dependencies which are used to calculate an order (\Fac) or the user specifies no ordering and search is used (\Stroll and some aspects of \Fac).

\subsection{Comparison to backward build systems}
\label{sec:remote_execution}

The design space of backward build systems is discussed in \cite{build_systems_a_la_carte}. In that paper it is notable that forward build systems do not naturally fit into the design space, lacking the features that a build system requires. We feel that omission points to an interesting gap in the landscape of build systems. We think that it is likely forward build systems could be characterised similarly, but that we have yet to develop the necessary variety of forward build systems to do so. There are two dimensions used to classify backward build systems:

\textbf{Ordering} Looking at the features of \Rattle, the ordering is a sequential list, representing an excessively strict ordering given by the user. The use of speculation is an attempt to weaken that ordering into one that is less linear and more precise.

\textbf{Rebuild} For rebuilding, \Rattle looks a lot like the constructive trace model -- traces are made, stored in a cloud, and available for future use. The one wrinkle is that a trace may be later invalidated if it turns out a hazard occurred (see \S\ref{sec:choices}).

There are also two  features available in some backward build systems that are relevant to \Rattle:

\textbf{Sandboxing} Some backward build systems (e.g. \Bazel \cite{bazel}) run processes in a sandbox, where access to files which weren't declared as dependencies are blocked -- ensuring dependencies are always sufficient. A consequence is that often it can be harder to write a \Bazel build system, requiring users to declare dependencies like \texttt{gcc} and system headers that are typically ignored. The sandbox doesn't prevent the reverse problem of too many dependencies.

\textbf{Remote Execution} Some build systems (e.g. \Bazel and \BuildXL \cite{buildxl}) allow running commands on a remote machine, usually with a much higher degree of parallelism than is available on the users machine. If \Rattle was able to leverage remote execution then speculative commands could be used to fill up the cloud cache, and \emph{not} involve local writes to disk, eliminating all speculative hazards -- a very attractive property. Remote execution in \Bazel sends all files alongside the command, but \Rattle doesn't know the files accessed in advance, which makes that model infeasible. Remote execution in \BuildXL sends the files it thinks the command will need, augments the file system to block if additional files are accessed, and sends further requests back to the initial machine -- which would fit nicely with \Rattle.

\subsection{Analysis of existing build systems}

We aren't the first to observe that existing build systems often have incorrect dependencies.  \citet{bezemer2017empirical} performed an analysis of the missing dependencies in \Make build scripts, finding over \emph{1.2 million unspecified dependencies} among four projects. To detect missing dependencies, \citet{detecting_incorrect_build_rules} introduced a concept called \emph{build fuzzing}, finding race-conditions and errors in 30 projects. It has also been shown that build maintenance requires as much as a 27\% overhead \cite{build_maintenance}, a substantial proportion of which is devoted to dependency management. Our anecdotes from \S\ref{sec:evaluation} all reinforce these messages.

\subsection{Speculation}

Speculation is used extensively in many other areas of computer science, from processors to distributed systems. If an action is taken before it is known to be required, that can increase performance, but undesired side-effects can occur. Most speculating systems attempt to block the side-effects from happening, or roll them back if they do.

The most common use of speculation in computer science is the CPU -- \citet{swanson_cpu_speculation} found that 93\% of useful CPU instructions were evaluated speculatively. CPUs use hazards to detect incorrect speculation, with similar types of read/write, write/write and write/read hazards \cite{patterson_cpu_design} -- our terminology is inspired by their approaches. For CPUs many solutions to hazards are available, e.g. stalling the pipeline if a hazard is detected in advance or the Tomasulo algorithm \cite{tomasulo}. We also stall the pipeline if we detect potential hazards, although have to do so with incomplete information, unlike a CPU. The Tomasulo algorithm involves writing results into temporary locations and the moving them over afterwards -- use of remote execution (\S\ref{sec:remote_execution}) might be a way to apply similar ideas to build systems.

Looking towards software systems, \citet{welc2005safe} showed how to add speculation to Java programs, by marking certain parts of the program as worth speculating with a future. Similar to our work, they wanted Java with speculation to respect the semantics of the sequential version of the program, which required two main techniques. First, all data accesses to shared state are tracked and recorded, and if a dependency violation occurs the offending code is revoked and restarted.  Second, shared state is versioned using a copy-on-write invariant to ensure threads write to their own copies, preventing a future from seeing its continuation's writes.

Thread level speculation \cite{steffan1998potential} is used by compilers to automatically parallelise programs, often by executing multiple iterations of a loop body simultaneously. As before, the goal is to maintain the semantics as single-threaded execution. Techniques commonly involve buffering speculative writes \cite{steffan2000scalable} and ensuring that a read reflects the speculative writes of threads that logically precede it.

Speculation has also been investigated for distributed systems. \citet{nightingale2005speculative} showed that adding speculation to distributed file systems such as NFS can make some benchmarks over 10 times faster, by allowing multiple file system operations to occur concurrently. A model allowing more distributed speculation, even in the presence of message passing between speculated distributed processes, is presented by \citet{tapus2006distributed}. Both these pieces of work involve modifying the Linux kernel with a custom file system to implement roll backs transparently.

All these approaches rely on the ability to trap writes, either placing them in a buffer and applying them later or rolling them back. Unfortunately, such facilities, while desirable, are currently difficult to achieve in portable cross-platform abstractions (\S\ref{sec:tracing}). We have used the ideas underlying speculative execution, but if the necessary facilities became available in future, it's possible we could follow the approaches more directly.

%% file: 7-conclusion.tex
\section{Conclusion and future work}
\label{sec:conclusion}

In this paper we present \Rattle, a build system that  takes a sequence of actions and treats them as a build script. From the user perspective, they get the benefits of a conventional build system (incrementality, parallelism) but at lower cost (less time thinking about dependencies, only needing to supply a valid ordering). Compared to conventional build systems like \Make,  \Rattle presents a simpler user interface (no dependencies), but a more complex implementation model.

Our evaluation in \S\ref{sec:evaluation} shows that for some popular real-world projects, switching to \Rattle would bring about simplicity and correctness benefits, with negligible performance cost. The two places where builds aren't roughly equivalent to \Make are the initial build (which could be solved with a global shared cache) and when speculation leads to a hazard. There are several approaches to improving speculation, including giving \Rattle a list of commands that should not be speculated (which can be a perfect list for non-monadic builds), or giving \Rattle a subset of commands' inputs/outputs (like \Fac does), or implementing better recovery strategies from speculation errors.

Our evaluation focuses on projects whose build times are measured in seconds or minutes, with only one project in the hours range. It is as yet unclear whether similar benefits could be achieved on larger code bases, and whether the \Rattle approach of ``any valid ordering'' is easy to describe compositionally for large projects.

Our next steps are scaling \Rattle and incorporating feedback from actual users.